\documentstyle[epsfig]{aprim10}
\input{epsf}

\newif\ifAMStwofonts

\ifoldfss
  \ifCUPmtlplainloaded \else
    \NewTextAlphabet{textbfit} {cmbxti10} {}
    \NewTextAlphabet{textbfss} {cmssbx10} {}
    \NewMathAlphabet{mathbfit} {cmbxti10} {} 
    \NewMathAlphabet{mathbfss} {cmssbx10} {} 
  \fi
  \ifAMStwofonts
    \ifCUPmtlplainloaded \else
      \NewSymbolFont{upmath} {eurm10}
      \NewSymbolFont{AMSa} {msam10}
      \NewMathSymbol{\upi}     {0}{upmath}{19}
      \NewMathSymbol{\umu}     {0}{upmath}{16}
      \NewMathSymbol{\upartial}{0}{upmath}{40}
      \NewMathSymbol{\leqslant}{3}{AMSa}{36}
      \NewMathSymbol{\geqslant}{3}{AMSa}{3E}

    \fi
  \fi
\fi 

\ifnfssone
  \newmathalphabet{\mathit}
  \addtoversion{normal}{\mathit}{cmr}{m}{it}
  \addtoversion{bold}{\mathit}{cmr}{bx}{it}
  \newmathalphabet{\mathbfit} 
  \addtoversion{normal}{\mathbfit}{cmr}{bx}{it}
  \addtoversion{bold}{\mathbfit}{cmr}{bx}{it}
  \newmathalphabet{\mathbfss} 
  \addtoversion{normal}{\mathbfss}{cmss}{bx}{n}
  \addtoversion{bold}{\mathbfss}{cmss}{bx}{n}
  \ifAMStwofonts
    \ifCUPmtlplainloaded \else
      \UseAMStwoboldmath
      \makeatletter
      \new@mathgroup\upmath@group
      \define@mathgroup\mv@normal\upmath@group{eur}{m}{n}
      \define@mathgroup\mv@bold\upmath@group{eur}{b}{n}
      \edef\UPM{\hexnumber\upmath@group}
      \new@mathgroup\amsa@group
      \define@mathgroup\mv@normal\amsa@group{msa}{m}{n}
      \define@mathgroup\mv@bold\amsa@group{msa}{m}{n}
      \edef\AMSa{\hexnumber\amsa@group}
      \makeatother
      \mathchardef\upi="0\UPM19
      \mathchardef\umu="0\UPM16
      \mathchardef\upartial="0\UPM40
      \mathchardef\leqslant="3\AMSa36
      \mathchardef\geqslant="3\AMSa3E
    \fi
  \fi
\fi 

\ifnfsstwo
  \DeclareMathAlphabet{\mathbfit}{OT1}{cmr}{bx}{it}
  \SetMathAlphabet\mathbfit{bold}{OT1}{cmr}{bx}{it}
  \DeclareMathAlphabet{\mathbfss}{OT1}{cmss}{bx}{n}
  \SetMathAlphabet\mathbfss{bold}{OT1}{cmss}{bx}{n}
  \ifAMStwofonts
    \ifCUPmtlplainloaded \else
      \DeclareSymbolFont{UPM}{U}{eur}{m}{n}
      \SetSymbolFont{UPM}{bold}{U}{eur}{b}{n}
      \DeclareSymbolFont{AMSa}{U}{msa}{m}{n}
      \DeclareMathSymbol{\upi}{0}{UPM}{"19}
      \DeclareMathSymbol{\umu}{0}{UPM}{"16}
      \DeclareMathSymbol{\upartial}{0}{UPM}{"40}
      \DeclareMathSymbol{\leqslant}{3}{AMSa}{"36}
      \DeclareMathSymbol{\geqslant}{3}{AMSa}{"3E}
    \fi
  \fi
\fi 

\ifCUPmtlplainloaded \else
  \ifAMStwofonts \else 
    \def\upi{\pi}
    \def\umu{\mu}
    \def\upartial{\partial}
  \fi
\fi

\title[]{New Views of EIT Wave and CME from STEREO}

\author[Ma et al.]
       {S. Ma$^{1, 2}$, J. Lin$^{1, 3}$, P. Chen$^4$ and H. Chen$^5$\\
        $^1$National Astronomical Observatories of
China/Yunnan Astronomical Observatory, \\Chinese Academy of
Sciences(CAS), Kunming, Yunnan 650011, China.\\
        $^2$Graduate School of CAS, Beijing 100049, China.\\
        $^3$Harvard-Smithsonian Center for Astrophysics, 60 Garden
        Street, Cambridge, MA 02138, USA\\
        $^4$Department of Astronomy, Nanjing University,
Nanjing, Jiangsu 210093, China\\
        $5$China University of Petroleum, 271 2nd North Street, Dongying, Shandong 257061,China}
\date{}

\pagerange{\pageref{firstpage}--\pageref{lastpage}} \pubyear{2008}
\begin{document}

\maketitle

\label{firstpage}

\begin{abstract}
On 2007 December 7, a small filament located in a small active
region AR 10977 erupted and led to a B1.4 flare. An EIT wave
associated with this eruption was observed both by SOHO/EIT and by
EUVI on board STEREO. According to the observations from
SOHO/LASCO and STEREO/COR A, we found that there was no CME
associated with the EIT wave. This seems to challenge the argument
that the cause of EIT waves is CME. However the data from
STEREO/COR B indicated that there was a narrow CME associated with
the EIT wave. This suggests that studying CMEs by investigating
observations made in one direction alone may not be able to
guarantee the reliability of the results.
\end{abstract}


\section{Introduction}
Usually, EIT waves appear as almost circular diffuse emission
enhancements propagating across the whole solar disk immediately
followed by an expanding dimming region if the magnetic structure
on the Sun is simple with only one active region on the disk
\cite{tho98}. While, when the global magnetic structure gets
complicated, they propagate rather inhomogeneously, avoiding
strong magnetic features and neutral lines and generally stopping
near coronal holes \cite{tho99}. Although EIT waves were found to
be a quite frequent phenomenon, there are still some questions
open. The cause of the EIT wave is still being debated between a
``flare-driven" and a ``CME-driven"
\cite{uch68,war01,war04,bie02,hud03,zhu04,cli05,vrs06}. The launch
of the Solar Terrestrial Relation Observatory (STEREO)
\cite{kai08} provides us an opportunity to find the answer.

\section{Instruments and Observations}

In this work we use the data from extreme ultraviolet imager
(EUVI), part of the Sun Earth Connection Coronal and Heliospheric
Investigation(SECCHI) \cite{how08} on board STEREO, which consists
of two identical spacecrafts that orbit the Sun ahead (STEREO A)
and behind (STEREO B) the Earth near the ecliptic plane. Their
separation angle was 42.4$\degr$ on 2007 December 7. EUVI images
the solar chromosphere and the low corona in four emission lines,
171 {\AA}, 195 {\AA}, 284 {\AA} and 304 {\AA}, out to 1.7
R$_\odot$. The pixel limited spatial resolution of EUVI images is
1.6$\arcsec$. In this research we focus on the EUVI images in 171
{\AA} and 195 {\AA} with a time cadence of 2.5 and 10 minutes
respectively.

COR1 is a classic Lyot internally occulting refractive
coronagraph. It images the inner corona with the pixel spatial
resolution of 3.75$\arcsec$ from 1.4 to 4 R$_\odot$ and the image
sequence cadence is 8 minutes. COR1 measures both the total
brightness B and the polarized brightness pB of the solar K
corona. In this paper we use the total brightness B. COR2 is an
outer coronagraph imaging the outer corona with the pixel spatial
resolution of 14.7$\arcsec$ and the image sequence cadence is 15
minutes from 2.5 to 15 R$_\odot$. The COR2 instruments acquires
only polarized images of the corona since the polarizer is always
in the beam. The total brightness B is used here.

\section{Results}

The EIT wave we studied occurred on 2007 December 7 during
4:25-5:10 UT, associated with the weak B1.4 class flare as well as
the eruptions of a filament and a sigmoid located in NOAA active
region AR 10977. On that day STEREO A was 20.8$\degr$ ahead the
Earth at a heliocentical distance of 0.97 AU and STEREO B was
21.6$\degr$ behind at 1.03AU.

The EIT wave showed a weak front in EUVI 171 {\AA} running
difference images (the top two rows of Figure 1) and a strong
front in EUVI 195 {\AA} ones (the bottom two rows of Figure 1).
The wave front of the EIT wave appeared different shapes in EUVI A
and B especially at the beginning of the EIT wave (see the panels
before 04:40 UT): the EIT wave looked like a right ear in EUVI A
and a left ear in EUVI B images. This indicates that the
propagation of the EIT wave at the beginning was not symmetric
about the center. However with the time progressing the EIT front
became more and more symmetric (see the panels after 04:40 UT).

Figure 2 displays the STEREO COR1 and COR2 white light images
showing the inner and outer corona intensity changes. It is very
interesting to notice that STEREO B observed the CME associated
with the EIT wave in both COR1 and COR2 FOVs, while STEREO A did
not. Nor did SOHO/LASCO. The second and the bottom rows display
the evolution of the CME. COR1 B showed that when the CME appeared
on the limb, its angular width is about 50$\degr$. Then its
angular width increased to about 90$\degr$. However, the angular
width of the same CME in COR2 images was about 28$\degr$ remained
roughly unchanged.

\begin{figure} 
 \centerline{{\epsfxsize=8.5cm\epsffile{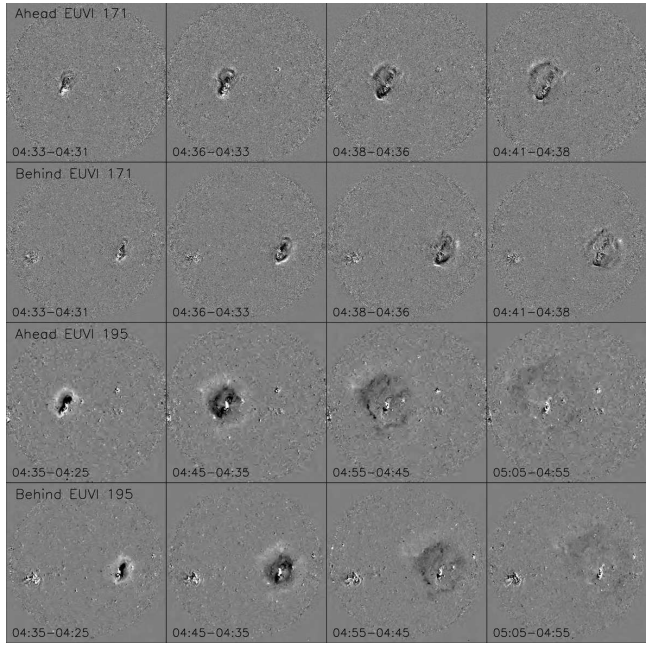}}}
 \caption[]{EUVI running difference
images showing the propagation of the EIT wave. The top two rows
for Ahead and Behind EUVI 171 {\AA} images and the bottom two rows
for Ahead and Behind EUVI 195 {\AA} images.}
\end{figure}

\begin{figure}
\centerline{{\epsfxsize=8.5cm\epsffile{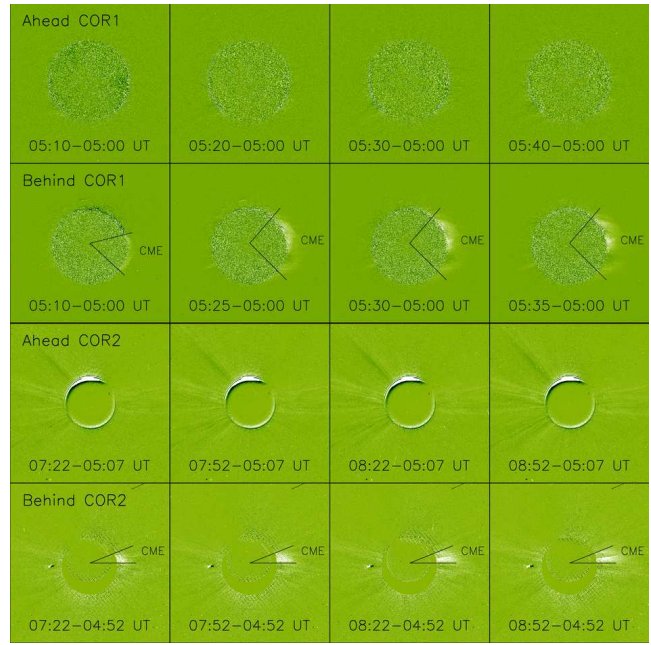}}}
\caption[]{STEREO COR1 (top two rows) and COR2 (bottom two rows)
different images displaying the CME.}
\end{figure}

\section{Conclusions and Discussions}
In this paper we displayed an observation of an EIT wave and the
associated CME in two directions simultaneously by STEREO A and B.
We found no CME associated with the EIT wave in the STEREO A data,
however a contrary result was obtained from STEREO B. Comparing
the data from STEREO A and that from STEREO B, we realized that
the CME was very narrow and roughly propagated toward STEREO A. So
it was almost totally blocked by the occulting disk of STEREO A at
the beginning, and it became too faint to be observed as its
angular width got large enough. This work reminds us of a preview
work by Thompson (2000) that reported an EIT wave without clear
evidence of a CME and suggested that some EIT waves may be
CME-poor. However, the present work indicates whether a CME is
observed to associate with an EIT wave depends on the angle at
which it is seen and on its brightness.

\section*{Acknowledgment}
We are grateful to the STEREO, EUVI, COR, SOHO, and LASCO teams
for their open data policy. This work was supported by MSTC grant
2006CB806303, by NSFC grants 10873030 and 40636031, and by CAS
grant KJCX2-YW-T04 to YNAO. JL was supported by NASA grant
NNX07AL72G when visiting CfA.

\label{lastpage}

\clearpage

\end{document}